\documentclass[a4paper,12pt]{article}

\usepackage[english]{babel}
\usepackage[utf8x]{inputenc}
\usepackage[T1]{fontenc}

\usepackage[a4paper,top=2.5cm,bottom=2cm,left=2cm,right=2cm,marginparwidth=1.75cm]{geometry}

\usepackage{amsmath}
\usepackage{amssymb}
\usepackage{amsthm}
\usepackage{graphicx}
\usepackage{comment}
\usepackage{authblk}
\usepackage{afterpage}
\usepackage{graphicx}
\usepackage[colorinlistoftodos]{todonotes}
\usepackage[colorlinks=true, allcolors=blue]{hyperref}

\title{ECONOMIC COMPLEXITY: “BUTTARLA IN CACIARA” 
VS A CONSTRUCTIVE APPROACH
}
\author[1,2]{Luciano Pietronero\thanks{luciano.pietronero@roma1.infn.it}}
\author[2]{Matthieu Cristelli}
\author[3]{Andrea Gabrielli}
\author[1]{Dario Mazzilli}
\author[2]{Emanuele Pugliese}
\author[2]{Andrea Tacchella}
\author[2]{Andrea Zaccaria}

\small{
\affil[1]{Dip. Di Fisica, Sapienza Univ. di Roma, P.le A.Moro 5, 00185, Roma}
\affil[2]{Istituto dei Sistemi Complessi (ISC)-CNR, UOS Sapienza, Dip. di Fisica, Sapienza Univ. di Roma, P.le A.Moro 5, 00185, Roma}
}

\newcommand{\be}{\begin{equation}}
\newcommand{\ee}{\end{equation}}
\newcommand{\bea}{\begin{eqnarray}}
\newcommand{\eea}{\end{eqnarray}}

\begin{document}
\maketitle
\abstract{This note is a contribution to the debate about the optimal algorithm for Economic Complexity that recently appeared in the Archive \cite{ECI+, hidalgo729} . The authors of \cite{hidalgo729} eventually agree that the ECI+ algorithm \cite{ECI+} consists just in a renaming of the Fitness algorithm we introduced in 2012 as we explicitly showed in \cite{ECI+answer}. However, they omit any comment on the fact that their extensive numerical tests claimed to demonstrate that the same algorithm works well if they name it ECI+, but not if its name is Fitness. They should realize that this eliminates any credibility to their numerical methods and therefore also to their new analysis, in which they consider many algorithms \cite{hidalgo729}. Since by their own admission the best algorithm is the Fitness one, their new claim became that the search for the best algorithm is pointless and all algorithms are alike. This is exactly the opposite of what they claimed a few days ago and it does not deserve much comments. After these clarifications we also present a constructive analysis of the status of Economic Complexity, its algorithms, its successes and its perspectives. For us the discussion closes here, we will not reply to further comments.}

\section*{"Buttarla in caciara"}
This is a picturesque expression of the roman slang that represents very well the situation of this discussion and has no appropriate translation in English. Thus we prefer to use it in the original language and, from the text, the reader will easily get the point.

A couple of weeks ago a paper appeared in arXiv \cite{ECI+} introducing a supposed new algorithm, ECI+, based on important new concepts. According to the authors, the algorithm was then tested extensively in various ways with respect to two other algorithms and the conclusion was that ECI+ represented by far the best algorithm for Economic Complexity.  In addition to the paper, this was announced as an important discovery also in conferences and social networks, e.g. on Facebook July 26: “I am glad to announce ECI+: a better Economic Complexity Index. ECI+ […] is more accurate and consistent when it comes to predict future economic growth.”\footnote{https://www.facebook.com/cesifoti/posts/10155124955931693}

Shortly after we submitted a note that showed that ECI+ is precisely identical to the Fitness algorithm we published in 2012 and on which we have done extensive work since then \cite{scirep2012,plos2013,sahara,cimini14, plos2015,netherlands,conv,growth,plos2017}. However, the big surprise was that the authors of \cite{ECI+} also presented a detailed and extensive numerical analysis finding the incredible result that the same algorithm, the Fitness, works very well with the name ECI+, but not if it is named Fitness. This clearly poses a problem of credibility for these numerical analyses. This is then confirmed by the inability of the authors to replicate the published results for the Fitness \cite{scirep2012, plos2013, plos2015}, which were not even mentioned in their paper.

In the subsequent addendum \cite{hidalgo729} they reluctantly must agree that the algorithm of ECI+ is just a renaming of the original Fitness. So one could now think that the issue is over and the best algorithm, at least for the moment, is the Fitness. Not at all. They now try to set the agenda in a completely different perspective and to confuse the whole discussion. Indeed they argue that most of the algorithms are basically all alike and the search for an optimal algorithm is actually pointless. However, precisely this search was the object of their former paper a few weeks ago. 

A crucial point is that no comment is made on the points we raised about the total inconsistency of their numerical analyses and their inability to reproduce the published rankings for the Fitness. In this respect one should note that the new claim that algorithms are all alike is based on the same type of analysis that showed that the same algorithm is more or less good depending on its name. For these reasons we consider this whole analysis as pointless.

\section*{Are really all algorithms alike?}
First of all there is a fairness problem. Few weeks ago these authors were arguing to have made the big discovery of the ECI+ algorithm, which was claimed to be the best one \cite{ECI+}. After we pointed out that ECI+ is just a renaming of the Fitness \cite{ECI+answer}, they try to teach us that the search for the optimal algorithm is actually pointless \cite{hidalgo729}. Flipping the table will not do any good to the level of the scientific discussion.

Algorithms correspond to the mathematical translation of basic scientific concepts. They permit to quantitatively test these concepts and to perform useful analyses and predictions. In this respect they are the analogous to physical laws for natural phenomena. Along this analogy and following their reasoning, one could argue that the basic experimental evidence for gravitation is that, if one throws a stone out of the window, this goes down and not up. It would be easy to check that 100 different gravitational theories would agree with this observation and then conclude that the search for the best theory is pointless. One could also argue that if the force is proportional to the velocity or to the acceleration is just a minor irrelevant point etc.

About network algorithmics, along the same reasoning, one could also have argued that all algorithms are alike, while, on the contrary, the Google PageRank ™ (GPR) turned out to be much better and outperformed all the others. This proves in a spectacular way that also in this field, actually close to our problems, the choice of the best algorithm is crucial.

Clearly the authors themselves cannot believe in the arguments of the irrelevancy of the identification of the optimal strategy as proven by their own article of a few weeks ago: As they stated in their abstract \cite{ECI+} “We find that ECI+ outperforms ECI and Fitness”, ECI being the output of an algorithm introduced by Hidalgo and Hausmann \cite{ECI}. Such a conclusion becomes embarrassing now that they had to admit that ECI+ is equal to Fitness. Therefore the new arguments that all algorithms are alike are introduced to avoid the more embarrassing admission about the total inconsistency of their numerical analysis, in addition to the confusion they made about the evident mathematical properties of the algorithms. This is a perfect example of “buttarla in caciara”.

\section*{Constructive Discussion: The Search for Algorithms
}
Once we have clarified that this is actually an important and interesting problem, we can turn to a constructive analysis of the algorithms of Economic Complexity.

Our starting point is given by the observational data: the matrix $M_{cp}$ that quantifies the ability of the country $c$ in exporting the product $p$. Operatively it is convenient to consider it as a binary matrix obtained by applying suitable thresholds to functions of the exports \cite{BACI} of each country in each product, but there is no problem to use continuous data and the results are essentially coherent. 

The problem is which is the optimal algorithm to define, from these observational data, the ranking of the industrial Fitness of countries ($F$) and the complexity of products ($Q$). We use our own notations for consistency in the whole discussion, referring to the Fitness of a country as the economic competitiveness of a country, defined as the output of a given algorithm.
It is easy to realize that the matrix $M_{cp}$ defines a bipartite country-product network. In this respect ECI \cite{ECI} defines the Fitness of a given country as the average Complexity of its products  and the Complexity of a product as the average Fitness of the countries which produce such a product. As such it was a reasonable attempt. However, things are not so simple, and actually much more interesting, to be captured by simple linear averages. 

\section*{The problems of the ECI algorithm}

The main problems of ECI (Economic Complexity Index, and PCI, Products Complexity Index, see \cite{ECI}) were of two types, conceptual and practical. We focus just on a few representative examples. In \cite{plos2013} one can find a more complete discussion.

\subsection*{Conceptual problems of ECI}

\begin{itemize}
\item{\textbf{The averaging:}
A simple visual inspection of the data of the matrix $M_{cp}$ clearly shows (see e.g. Fig. 1 in \cite{scirep2012}) a nested structure of the countries-products system, which implies a great importance of diversification (rather than specialization) in the export basket of a country. However, in the ECI algorithm diversification has absolutely zero importance. This has the paradoxical effect that if country $A$ exports 10 products whose Complexity goes from 1 to 10 and country $B$ exports only one product with Complexity equal to 6, the ECI algorithm would assign a fitness value $F(A)=5.5$ and $F(B)=6$. Consequently, following ECI, a country which makes many products, also of very high complexity, and including the one with $Q=6$, is ranked below the one that produces only this product (Fig.\ref{example}). Clearly this is a major conceptual problem, which, as we show below, produces totally unrealistic results.
\begin{figure}[htbp]
\centering
\includegraphics{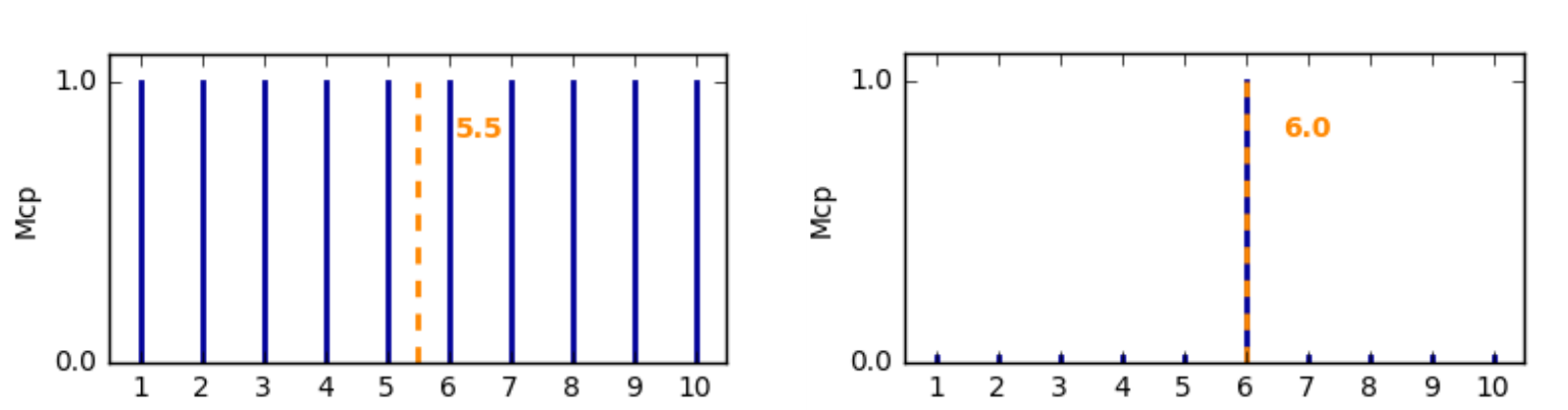}
\caption{Example of a two-countries model: Country $A$ (left) exports 10 products whose Complexity values range from 1 to 10, while country $B$ (right) exports a single product of Complexity 6. According to ECI the Fitness of $A$ is less than that of $B$, which is clearly a crucial inconsistency in the light of the nested structure of the matrix $M_{cp}$. This is due to the fact that ECI is based only on averages and diversification has zero effect.
}
\label{example}
\end{figure}
}
\item{\textbf{Linearity:}
The linearity of the relation between the Complexity of a product and the average Fitness of the countries that are able to produce it, may appear reasonable, but it leads to fundamental inconsistencies. For example consider Crude Oil (or in general any raw material of simple extraction). The availability of Oil is just a matter of luck and has nothing to do with the industrial competitiveness of a country. However, due to the linear arithmetic mean of ECI relating the Complexity of a product to the Fitnesses of its exporters, the high values of Fitness of countries like USA, UK, Norway and others pushes the Complexity of Oil up in a totally unrealistic way.
}
\end{itemize}

\subsection*{Problematic Results of ECI}
\begin{figure}[htbp]
\centering
\includegraphics[scale=0.8]{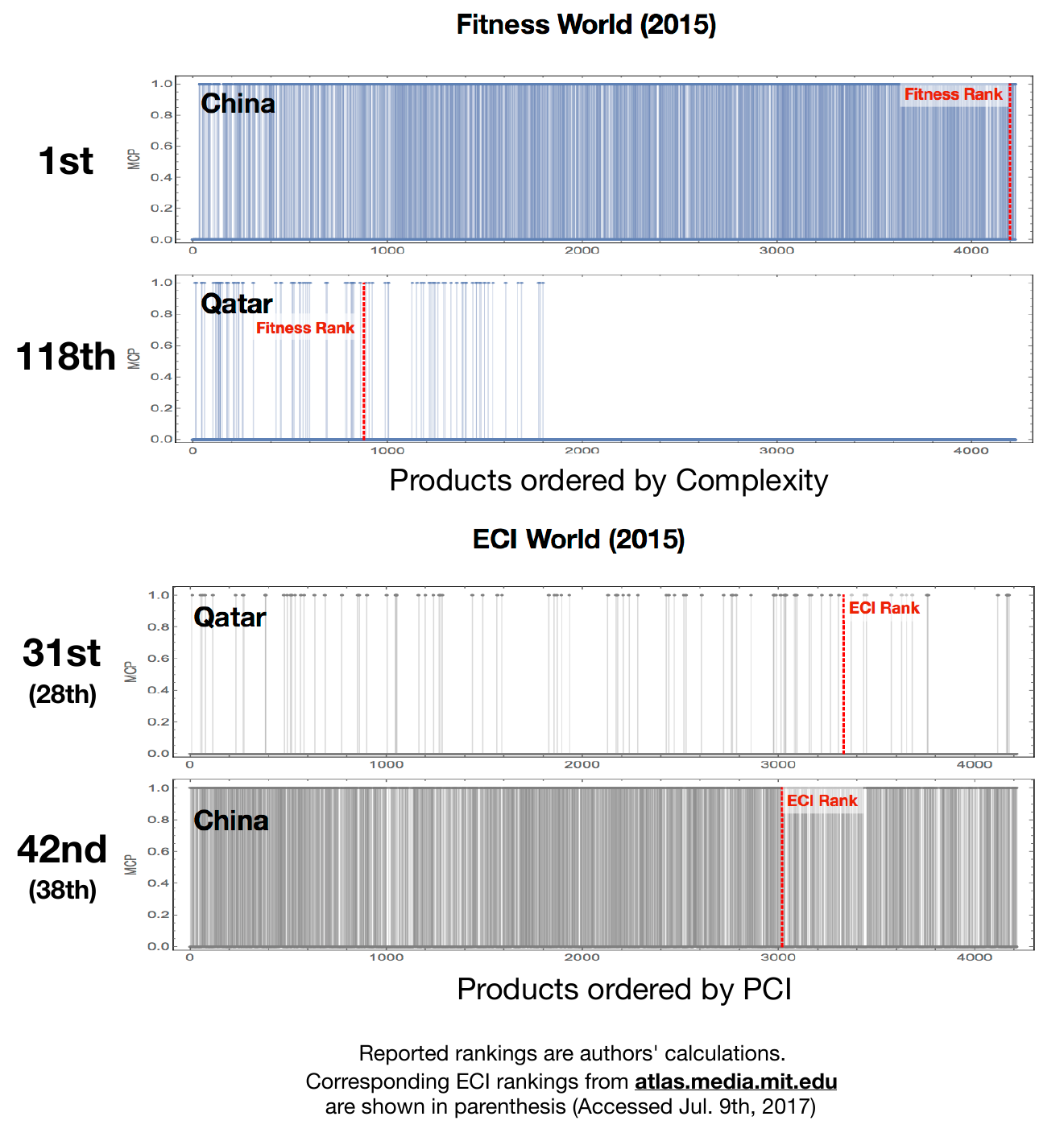}
\caption{Country Spectroscopy for Fitness and ECI, comparison between China and Qatar in 2015. Each bar represents a product in which the country is competitive, ordered by Complexity (upper panel) from Fitness algorithm and PCI (lower panel) from ECI algorithm. Red bars represent Fitness ranks (upper panel) and ECI rank (lower panel). Qatar produces essentially only Oil and rather simple related products. This is correctly reproduced by the Fitness algorithm that ranks Qatar very low and China very high. Instead ECI ranks these Oil products relatively high for the reasons we have discussed (linearity problem). In addition ECI defines the Fitness only as an average (average problem) and this leads to the paradoxical result that the ECI industrial competitiveness of China is below Qatar. 
}
\label{cq}
\end{figure}

The combination of these two conceptual and fundamental problems leads to incredibly unrealistic results like those presented in Fig.\ref{cq}. According to our reconstruction of the ECI ranking, and analogously to the rankings published by one of the authors of \cite{ECI+} in \cite{webarchive}, in 2015 Qatar ranks 31st in the world, while China is only 42nd. 

Actually, China has been the fastest growing economy in the world for the past 30 years and has reached a total GDP comparable to Europe and the US, while Qatar is a typical Oil country that produces essentially only that. Who can believe that Economic Complexity can be of any use if it gives such an absurd result? 
We are going to see that the Fitness algorithm gives completely different and much more reasonable results as shown in Fig.\ref{cq}. How such a fantastic discrepancy can be reconciled with the newest claim that algorithms are all the same is an exercise that we leave to the authors of \cite{hidalgo729}.

Here there can have been two kinds of problems or actually both at the same time. First, just mistakes like the ones that have been made in the erroneous reconstruction of the Fitness data \cite{ECI+}. In addition there can be a problem of intrinsic validity of the test. If a test shows that algorithms that give so different results are statistically similar with respect to some criterion, this simply means that such a criterion is inappropriate to discriminate their performances.

These are the kind of problems which led us to search for a conceptually different algorithm which could properly address these major inconsistencies and not just looking for some little improvement by playing randomly with parameters or exponents as done in \cite{hidalgo729}.
Our overall objective in this respect is to make Economic Complexity a realistic workable tool, better that the standard methods, and this cannot be done by putting these problems under the rug.
Note that our results shown for ECI in Fig.\ref{cq} were in strong agreement with those published in atlas.media.mit.edu until at least the day $9^{th}$ of July 2017, i.e. the last time we checked prior to writing this addendum. These are reported between parentheses in Fig.\ref{cq}. However, at the moment these rankings appear to have been modified in their website. Since we live in an era where information grows \cite{infogrows} but it cannot be destroyed, the previous rankings are still publicly available at \cite{webarchive} in a snapshot dated April 2017, which is identical to the page we accessed in July. 

We chose to refer to these previously published rankings for three reasons: 
\begin{enumerate}
\item These rankings are the ones the authors of \cite{ECI+} seem to refer to when they make comments about the differences between ECI and ECI+ such as “ECI+ and ECI have a strong correlation ($R^2$=85\%), but ECI+ tends to rank manufacturing heavy countries higher than ECI (ECI+ ranks Vietnam higher than Qatar and China higher than Norway)”. That wouldn’t be relevant anymore in light of the new rankings, since now China seems to be higher than Norway and Vietnam higher than Qatar for ECI as well; 
\item We are able to easily and almost exactly reproduce the previous rankings, but not the new ones, as demonstrated in Fig.\ref{cq}; 
\item The webpage displaying the ECI rankings has been modified recently, surely after July 9th 2017, without any comment on how such new results have been obtained or why 20 countries have been excluded from the new rankings.
\end{enumerate}
Point ii) in particular shows that we can properly reconstruct the ECI original results, contrary to the authors of \cite{ECI+} which have given totally wrong results for their Fitness reconstruction and did not even bother to consider that these are very different from those we originally published, more than 4 years ago \cite{plos2013}.

It should be pointed out that, anyway, similar striking results can be found even in the new ECI rankings, where, for instance, Slovakia is ranked as the 16th more complex economy in the world, above China, the Netherlands are 26th and India is 46th.

\subsection*{The origin of the Fitness Algorithm}
Starting from the above considerations about the problems of the averaging procedure in ECI, it is clear that we should have an algorithm that includes explicitly the diversification as an essential concept, so it is natural to avoid the average and consider only the sum over all products in the first equation. But this alone would miss the important information on the Complexity of the products so we weight the terms in this sum by the complexity of each product and get the first equation for the Fitness, which is still linear.
\[
{F}_c^{(n)}=\sum_p M_{cp} Q_{p}^{(n-1)}
\]
As for the Complexity of the products the situation is a bit more subtle. The first zero order consideration arising from the nested structure of the matrix $M_{cp}$ is that, in general, the more are the countries that can make a product, the less this product is complex. So it is natural to start by putting at the denominator the sum over all the countries that can make this product. However, this would consider all the countries alike, while the conceptual implications of the nested structure of $M_{cp}$ are deeper. For instance, the information that a product is made by a highly competitive country (high Fitness) like Germany should not decrease the complexity of the product itself as Germany makes essentially almost all products. But, if we know that the same product is exported by some countries with low Fitness, this fact is highly informative because such countries can make in general only few low Complexity products. This therefore determines a low Complexity for the considered product. From these considerations it is then natural to weight each term in the denominator with the inverse of the Fitness of each country.
\[
{Q}_p^{(n)}=\frac{1}{\sum_c M_{cp} \frac{1}{F_{c}^{(n-1)}}}
\]
In this way the lowest Fitness countries that can make the product give the maximum handicap to the value of its Complexity. This leads to the interesting fact that the natural algorithm for Economic Complexity is necessarily non-linear and therefore qualitatively different from both GPR and ECI. 

We can now see that the search of the optimal algorithm should be guided by careful mathematical arguments on the fundamental observational facts about the system (in our case the structure of the matrix $M_{cp}$) and not by a random check with variable parameters or exponents \cite{hidalgo729}\footnote{We note in passing that an analysis of the effect of adding exponents to the algorithm was already present in \cite{conv}, but the focus was the convergence properties of the algorithm.}. This example also shows that, contrary to the common wisdom that also led to ECI, the Google-like algorithms are not of general value for each problem in Big Data. One should study carefully the optimal algorithm on the basis of the general global structure of the data, which is a good news because it requires an interesting exercise of rational analysis and creativity, and many interesting developments and surprises can be expected in various different fields.

In summary, the ECI algorithm was just a simple linear average of local properties without any critical analysis of the implications of the global structure of the observational data (i.e. the matrix $M_{cp}$) and any consideration on the mathematical and conceptual problems that are present in the application of the ECI algorithm to these data \cite{plos2013}. In no case it was mentioned that it could be improved in any way, even in front of the absurd results it gives, like in the case of Qatar and China.

The Fitness algorithm instead arises from a careful analysis of these basic problems and from a logic reasoning for the identification of the key elements appropriate for Economic Complexity. It is a completely different perspective with respect to ECI and, even from a purely mathematical point of view, its distance from ECI is conceptual and fundamental, paving the way of similar reasoning in other fields. For example the Fitness algorithm was shown to be by far the best in the study of the stability of mutualistic ecosystems, which also have a nested structure \cite{munoz}.

We can now go back to the puzzling problem of China and Qatar and things will become immediately very clear and realistic. In Fig.\ref{cq} we can see that the Complexity of Qatar products (essentially Oil) is pushed up artificially by the ECI algorithm, which linearly relates the Complexity of Oil also to the Fitness of US and UK. Instead, in the nonlinear Fitness algorithm, the denominator of the Complexity formula is dominated by the countries with the lowest Fitness, and the role of US and UK becomes therefore irrelevant. This implies that in the Fitness algorithm the Complexity of Qatar products remains at a rather low value, as it is natural. The problem of the averaging is also crucial to resolve the puzzle. In fact one can see in Fig. 2 that for the ECI algorithm the Complexity of Qatar’s products are distributed also towards high values so, as in this approach their average leads to the country Fitness, this results superior to China. 

In conclusion the Fitness algorithm brings China to the first position (ranking of 2015) for industrial competitiveness (note that these data are mostly for manufacturing) and Qatar becomes 118th, a much more realistic situation with respect to the totally unrealistic respective values, 42nd and 31st, obtained with ECI.

We believe that this discussion of China and Qatar is a paradigmatic example that shows that the search of the proper algorithm requires a careful analysis of the problems and the implementation of new concepts within a coherent mathematical framework. The reader interested to more on this with tests on toy models and other mathematical discussions can found them in \cite{plos2013}.

So is the Fitness algorithm perfect? Not at all, but it is a solid starting point based on clear and understandable concepts and it overcomes the conceptual and mathematical inconsistencies of the ECI algorithm with a clear understanding of their origin.

Could one modulate slightly the exponents of the Fitness algorithm to better fit the data? In principle yes, but certainly not in the total confusion discussed in \cite{hidalgo729}. However, even if done properly we would not consider this particularly useful, at least at the moment.
The Fitness approach consists in a simple algorithm, based on clear concepts, which produces a global picture of the world industrial economy (mostly manufacturing for the moment but services are being added). However, this picture is \textit{impressionistic} in the sense that it is globally correct, but one cannot take a little portion and expand it with a microscope. If one is interested in the details one has to complement the Fitness with other data and different information and analysis as we are going to show in the following. A super optimized Fitness 2.0 algorithm, obtained playing only with parameters and exponents, would not help in any case to this purpose. In summary, the Fitness is a sound and reasonable starting point for a number of further developments, some of which we briefly mention below.

\section*{The Selective Predictability Scheme for Growth Forecasting}

\begin{figure}[htbp]
\centering
\includegraphics[scale=0.8]{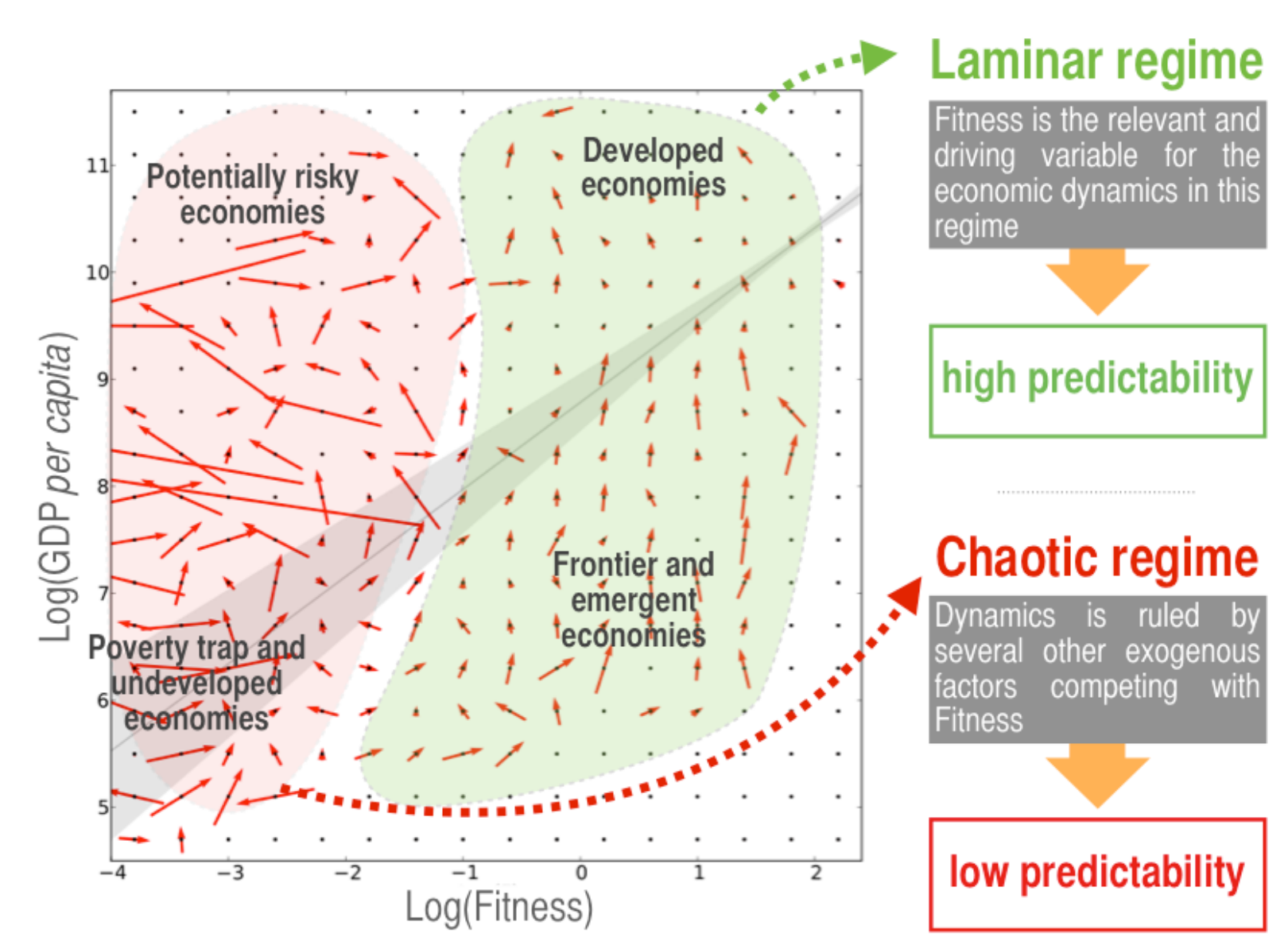}
\caption{Flow chart of the evolution of countries in the GDP and Fitness plane for the period 1995-2010. In the green zone we observe a laminar regime with high predictability while the red zone shows a chaotic behavior with low predictability.
}
\label{plane}
\end{figure}

The Fitness method we have developed synthesizes all the information about the industrial competitiveness of a country in a single variable: the Fitness. The idea is that the comparison of this intrinsic competitiveness with the GDP can reveal the \textit{hidden potential} for the development of a country. In Fig.\ref{plane} we can see a flow chart constructed from the time trajectories of all countries for the past 15 years. An interesting structure emerges: in the region of high Fitness (green) we observe a smooth, laminar behaviour of the flow. In the low Fitness region (red) instead we have a rather irregular and chaotic dynamics. 

The basic idea is that this is an optimal scheme to learn from the past. If all the countries that have passed in a certain region have all moved in the same direction in their future (laminar region), it is very likely that if a new country passes in this same region, it will follow the same path. This is the method of analogues \cite{analogues,plos2015}.
Clearly this approach leads to a Selective Predictability Scheme in which not only we can give a forecasting criterion, but we can also assign a degree of confidence to this prediction: high confidence in the laminar regime and low in the chaotic one. Note that this methodology is very different from the standard regressions that would mix the two zones in a single global fit. This method of local forecasting, which qualitatively reminds what is done for the weather forecasting, has been introduced in \cite{plos2015}.
The methods of analogous allows to make forecasting without any assumption on the functional form of the relationship between variables which is instead the basis of standard parametric regression. This is the same philosophy behind the very successful machine learning approaches.

\section*{Comparison with the state of the art (IMF)
}
In Ref. \cite{growth} we have performed a detailed study of the country GDP 5 years growth forecasting based on the Fitness method and we compared the results with the standard IMF analysis. The standard concepts to evaluate forecasting are:

MAE: Mean Absolute Error, which is the average of the absolute difference between forecasting and actual growth rate.
RMSE: Root Mean Squared Errors, which gives greater importance to large errors.

Note that these are “out of sample” tests while the classical correlations shown for regressions would be “in sample”.

\begin{figure}[htbp]
\centering
\includegraphics[scale=0.9]{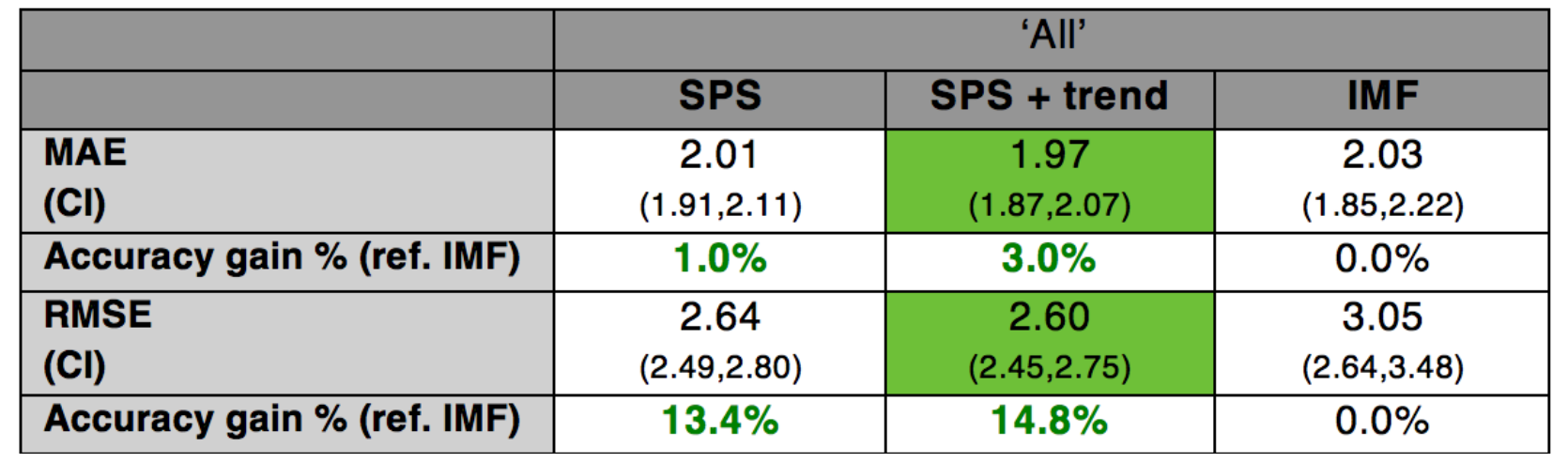}
\caption{Comparison of the performance of the forecasting of the Fitness approach with the standard IMF results. We report also the confidence intervals (CI), computed at 3 $\sigma$. See text for details.}
\label{tableall}
\end{figure}

In Fig.\ref{tableall} we report this analysis and comparison for all the countries. The Fitness method can be implemented in two ways, named SPS and SPS+trend. In the second one a degree of autocorrelation is also considered \cite{growth}.

An interesting point is that the Fitness forecasting is comparable, and slightly better, with respect to the IMF. However, it is important to note that in the Fitness approach the data used are much less than in IMF. In addition the algorithm and the method is precisely the same for all countries and such results can be easily reproduced by everybody. IMF instead adopts different criteria and a different team for each country and these data are then homogenized by a central team. In addition there are also a number of country specific assumptions.
The present analysis shows the power of the new Fitness analysis forecasting that, given its simplicity and its uniform strategy can probably be improved easily, possibly with an integration of elements from the IMF methodology.

\begin{figure}[htbp]
\centering
\includegraphics[scale=1]{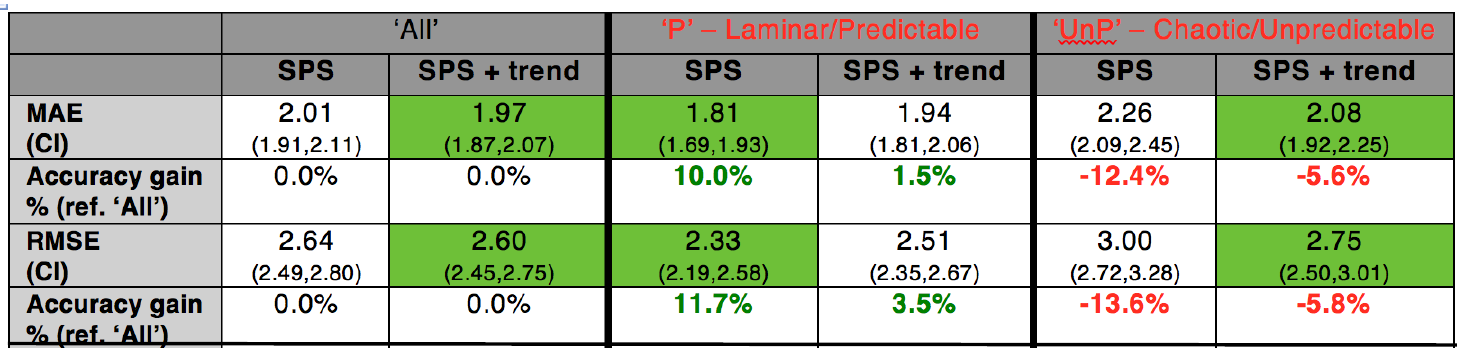}
\caption{Comparison of the forecasting approach of the Fitness method for the laminar and turbulent regions.
}
\label{tablepred}
\end{figure}

In Fig.\ref{tablepred} we show how forecasting is affected by the zone considered, laminar or chaotic. One can see that especially for RMSE the difference is appreciable. This is a very important point because it permits to introduce a novel concept that measures the \textit{predictability} of economic growth which is exclusive of the Fitness approach. Interestingly also the IMF results are better in correspondence of our areas of high predictability. This suggests that the concept of predictability is general and not related to our specific approach.

\subsection*{The mystery of China Growth: the big debate
}
\begin{figure}[htbp]
\centering
\includegraphics[scale=0.8]{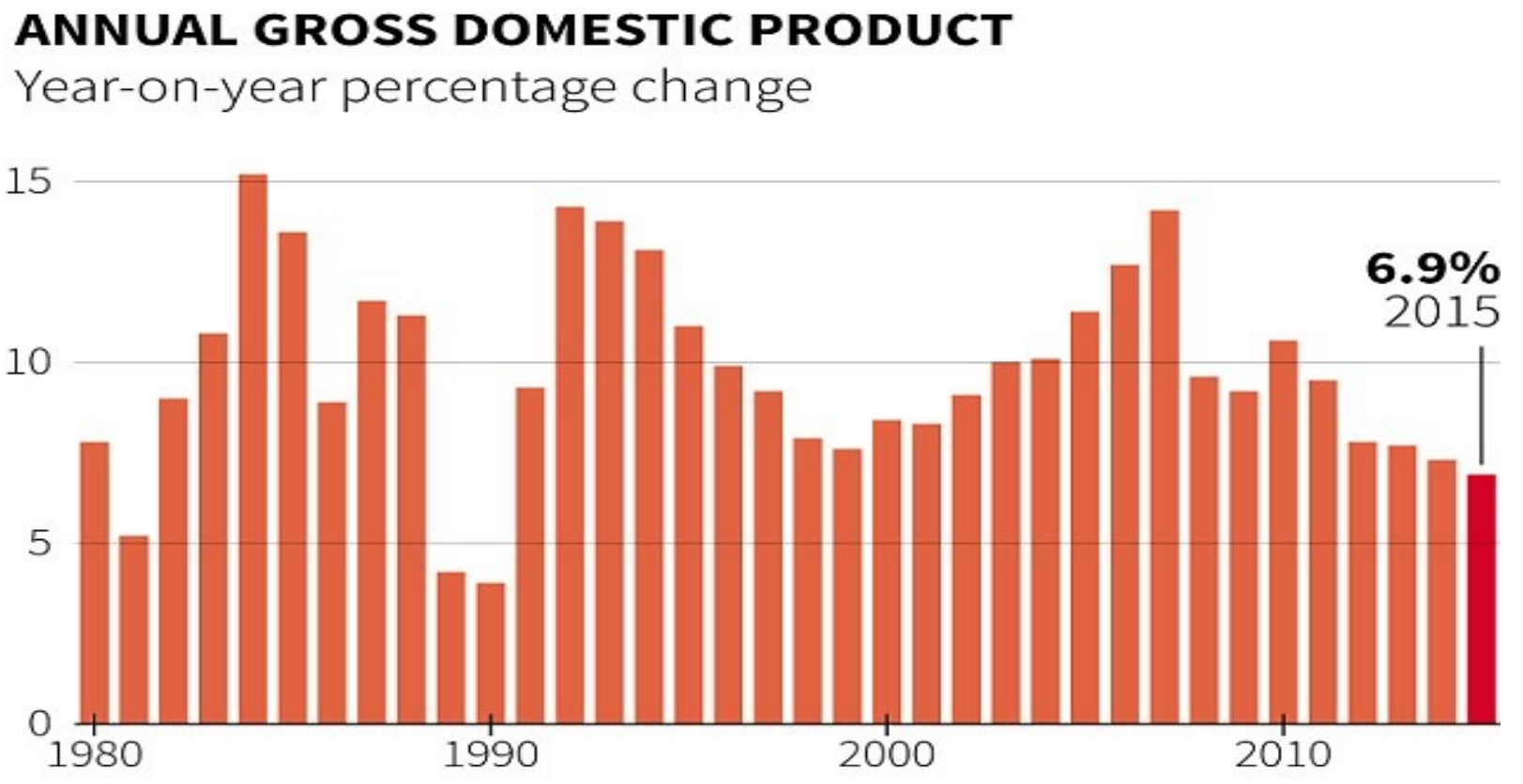}
\caption{GDP growth rate of China from 1980 to now
}
\label{china}
\end{figure}

The fantastic economic growth of China in the past 35 years (Fig.\ref{china}) represents an extremely interesting example with great implications and challenges for economic theory and political planning. Is China some sort of magic outlier or can it be casted in some rational conceptual scheme? The debate is hot.

Godfree Robert collected a huge number of opinions from authoritative sources since 1990 till today\footnote{See for instance https://www.quora.com/What-will-happen-if-Chinas-economic-growth-slows-to-3}. These claims predicted a hard landing of China growth almost every year from 1990 to now, but today China GDPpc is still growing at a rate of 6.9\% well above the present expectations. The problem is that China has had a growth of more than 5\% (sometimes much more) for the past 35 years. This is an absolute statistical anomaly which is hard to reconcile with the standard economic and statistical criteria.

For example in the Financial Times of Nov. 19, 2014 an editorial by David Pilling claims that 
“What goes up must come down – even China”. This article is a summary of an influential paper by Lant Pritchett and Lawrence Summers (PS, Ref.\cite{summers}) who argue that the single most robust and striking fact about growth is “regression to the mean” of about 2 per cent.  Only rarely in modern history, they say, have countries grown at “super-rapid” rates above 6\% for much more than a decade. China has managed to buck the trend since 1977 by harnessing market forces, engineering possibly the longest spell “in the history of mankind”. But what goes up, the authors tell us, must eventually come down. The argument of PS is based on the following statistical concepts:
The normal mean growth rate for industrial countries is between 0\% and 2\%. The periods of fast growth, above 5\% last a limited time of the order of ten years. For example in the 80’s Brazil had a fast growth above 5\% for 9 years but then, abruptly it went to zero growth and it stayed there for the following twenty years, now it is even negative. The regression to the mean is abrupt and very difficult to predict from the economic data of the country in question.
Considering that the period of fast growth for China is way out with respect to the standard duration we should expect, according to PS, that a regression to the mean could occur at any time, with potentially devastating consequences for the whole world economy.

\begin{figure}[htbp]
\includegraphics[scale=0.45]{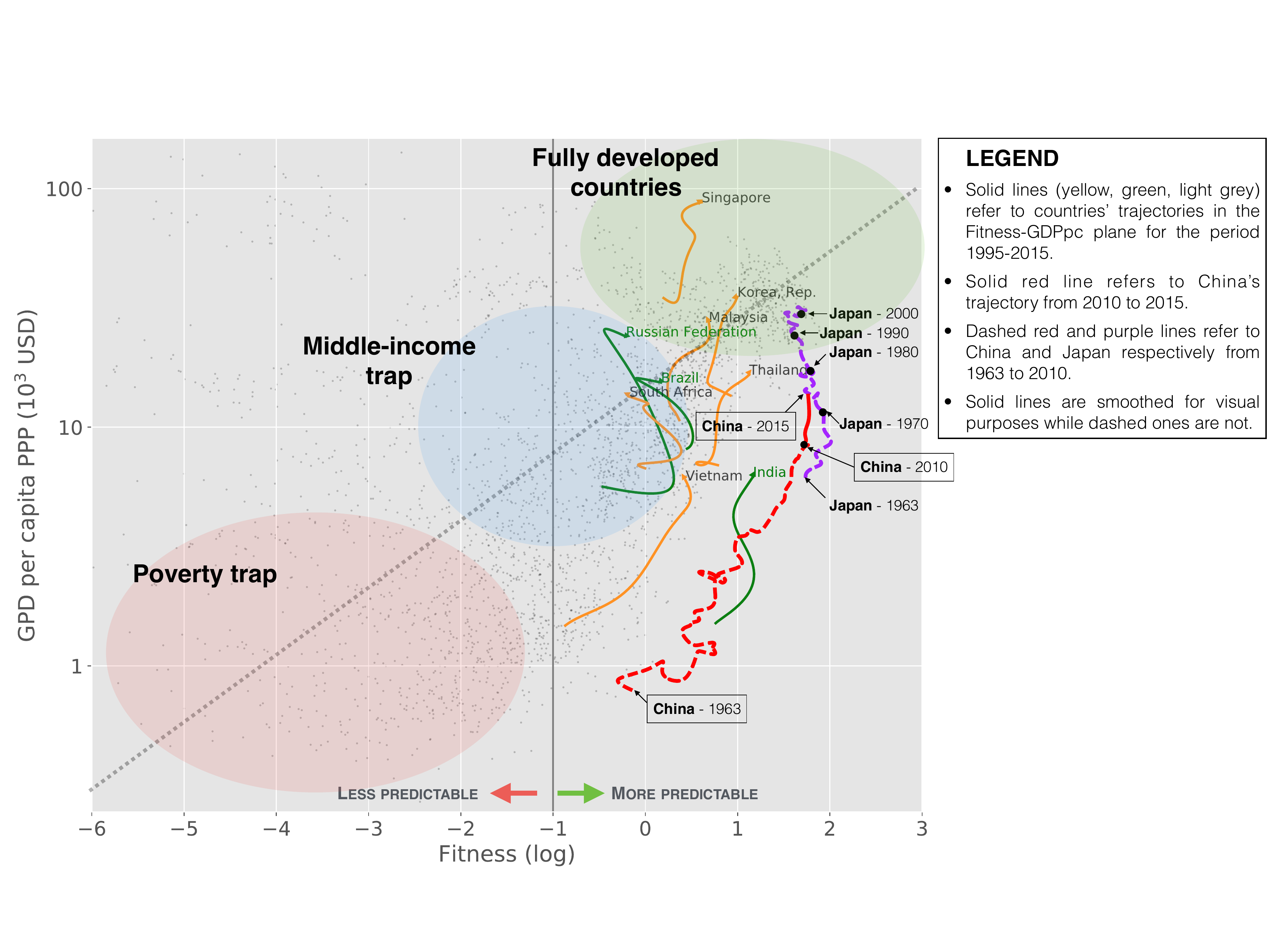}
\caption{Trajectories of various countries in the GDPpc-Fitness plane. One can see that, even if the GDPpc of Brasil and China are similar, the Fitness of China is far superior. This broader view permits us to understand why China has been growing so much for such a long period and also to forecast that further growth can be expected.
}
\label{chinatraj}
\end{figure}

The Fitness approach permits us to be much more optimistic. The PS argument is purely statistical and, within these limits, it is fine. But the Fitness permits an interpretation of these data with a different conclusion. We can see in Fig. \ref{chinatraj} the trajectories of various countries in the GDP-Fitness plane. The diagonal line represents a sort of equilibrium situation in which the GDPpc is in equilibrium with the Fitness. If a country is far below this line it means that its Fitness implies capabilities superior to the GDPpc actually realized. In such a situation we expect the country to have a hidden potential that can lead to fast growth and this fast growth stops when the trajectory reaches the diagonal line. This gives an interpretation of the abrupt stop of the fast growth as the reaching of an equilibrium between the country in question and the rest of the world. In this perspective it is not surprising that if one considers only the internal data of one individual country, this phenomenon is not visible.

Then we can see that even if Brazil and China have a similar GDPpc, the Fitness of China is much greater and the two trajectories have little in common. In this respect the two trajectories are totally different and Brazil cannot be taken as a possible example for China. This also means that China is already way out from the risk of falling in the Middle Income Trap were Brazil is instead located.
Finally, we can see that the trajectory of China is still far from the diagonal equilibrium line and this leads us to forecast that, from the point of view of industrial competitiveness, we can expect still some years of fast growth for China. In case a possible comparison can be made with Japan at a previous timescale.

\section*{Further developments}
These methodologies have been applied to the study of single countries \cite{netherlands} or geographical regions \cite{sahara}, as well as to the discussion of the possible strategies to escape from the poverty trap \cite{plos2017}.
Moreover the same framework, exploiting the nestedness and diversification of human activities in different fields, was used to understand the relationship between economic, scientific \cite{cimini14,cimini16} and technological activities. By matching data on the economic and innovation system, we linked patenting in each technological field with the development of a comparative advantage in a specific product: countries with a technological advantage in one field will be more likely, in the future, to successfully export one product. The same was done with scientific articles, to gain a novel understanding of the Global Innovation System linking science, technology and economic activities \cite{innovation}.
Not only this map is of great academic interest for the economics of innovation and development: it also allows for many practical exercises for the long-term strategic planning of countries and firms.

\begin{figure}[htbp]
  \centering
  \includegraphics[scale=0.4]{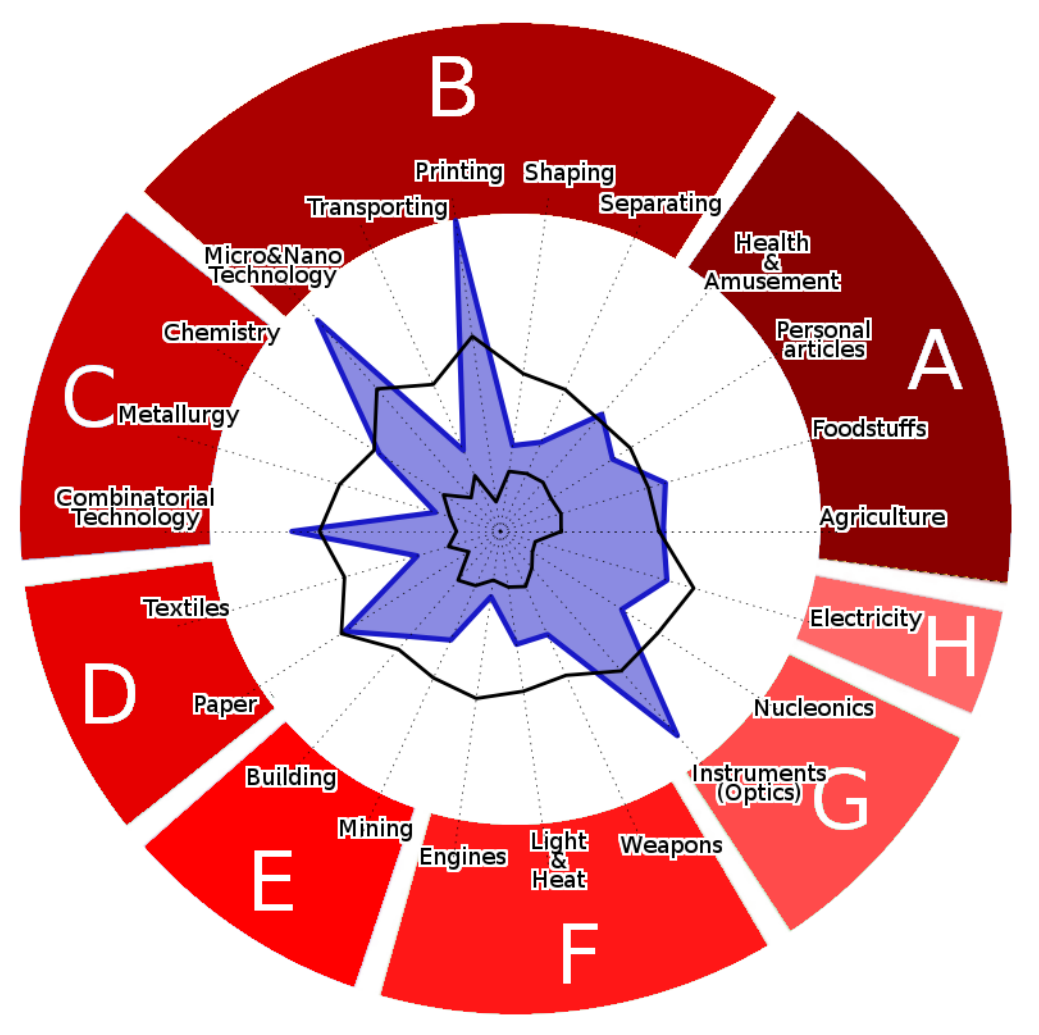}%
  \qquad\qquad
  \includegraphics[scale=0.6]{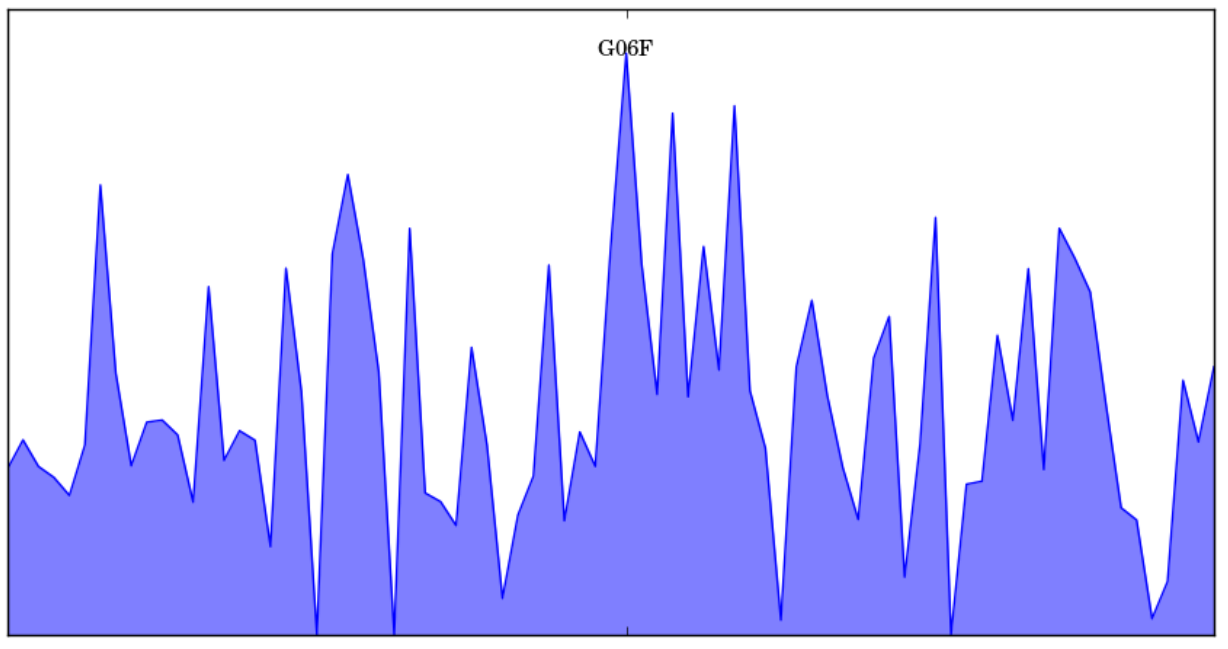}
  \caption{ Technological fields related to the future export of Desktop Computers. On the left we show the broad fields. On the right we further analyze the Optical field to show which are the main sub fields interested. The peak in the subclass G06F correspond to electrical digital data processing.}
\end{figure}

\section*{To find out more}
The interested reader can find more information about these methodologies on the website \url{www.economic-fitness.com}.\\
A compact summary of our activities can be found in this brochure: \url{https://www.economic-fitness.com/assets/media/ec4u.pdf}.

\end{document}